 \documentclass[aps,amssymb,prl,twocolumn,showpacs]{revtex4} \fussy
\usepackage{times}

\usepackage{graphicx} 

\newcommand{\figwidth}{0.45\textwidth} 

\begin{document}

\title{Electric and magnetic dipole coupling in near-infrared split ring metamaterial arrays}

\author{Ivana
Sersic}\email{i.sersic@amolf.nl}
\affiliation{Center for Nanophotonics, FOM Institute for Atomic and
Molecular Physics (AMOLF), Science Park 102, 1098 XG Amsterdam, The
Netherlands}
\author{Martin Frimmer}
\affiliation{Center for Nanophotonics, FOM Institute for Atomic and
Molecular Physics (AMOLF), Science Park 102, 1098 XG Amsterdam, The
Netherlands}
\author{Ewold Verhagen}
\affiliation{Center for Nanophotonics, FOM Institute for Atomic and
Molecular Physics (AMOLF), Science Park 102, 1098 XG Amsterdam, The
Netherlands}
\author{A. Femius Koenderink}
\affiliation{Center for Nanophotonics, FOM Institute for Atomic and
Molecular Physics (AMOLF), Science Park 102, 1098 XG Amsterdam, The
Netherlands}

\begin{abstract}
We present experimental observations of  strong electric and
magnetic interactions between split ring resonators (SRRs) in
metamaterials. We fabricated near-infrared (1.4~$\mu$m) planar
metamaterials with different inter-SRR spacings along different
directions. Our transmission measurements show blueshifts and
redshifts of the magnetic resonance, depending on SRR orientation
relative to the lattice. The shifts agree well with a simple model
with simultaneous magnetic and electric near-field dipole coupling.
We also find large broadening of the resonance, accompanied by a
decrease in effective cross section per SRR with increasing density.
These effects result from superradiant scattering. Our data shed new
light on Lorentz-Lorenz approaches to metamaterials.
\end{abstract}

\date{Submitted to Phys. Rev. Lett., July 15, 2009, resubmitted~October 1, 2009.}
\maketitle 

Since the seminal work of Veselago and Pendry~\cite{Pendry00}, many
experimentalists have started to pursue optical materials with
negative permittivity $\epsilon$ and permeability
$\mu$~\cite{Pendry01}. The key motivation is the prospect of
`transformation optics', which allows arbitrary bending of
electromagnetic fields, provided one has full control over
$\epsilon$ and $\mu$. Particularly exciting examples are perfect
lenses, that allow perfect sub-diffraction focusing~\cite{Pendry00},
and `cloaks' in which light passes an object without
scattering~\cite{Pendry06}. Full control over $\epsilon$ and $\mu$
requires `metamaterials' of artificial nano-scatterers with electric
and magnetic response, arranged in sub-wavelength arrays. The
archetypical  building block is the split ring resonator (SRR)
consisting of a single cut metal loop with an inductive response. In
recent years the field of metamaterials has made tremendous progress
in shifting the resonant response  from microwave  to optical
frequencies~\cite{Smith00, Enkrich05,Rockstuhl06,Shalaev05,Klein06}.
An important conceptual question is whether the effective response
captured by $\epsilon$ and $\mu$ is influenced by coupling between
constituents. Coupling between SRRs in vertical 1D
stacks~\cite{Shamonina02,Liu09a} has attracted great attention
lately outside the scope of metamaterials, \emph{e.g.}, for magnetic
waveguides~\cite{Brongersma00,Shamonina02,Koenderink06,Koenderink07},
antennas~\cite{Li06}, metamaterial lasers~\cite{Zheludev08}, and
stereomaterials~\cite{Zhang09}. Although constituent coupling might
be anticipated   to affect effective medium
parameters~\cite{Marques08}, measured effective responses have been
 attributed to single constituents in all experiments on  metamaterial arrays to date.

 In this Letter we
present the first measurements  of strong constituent coupling in
planar SRR metamaterial arrays. We fabricated and characterized SRR
lattices with a magnetic response at
$\lambda=1.4~\mu$m~\cite{Enkrich05,Rockstuhl06}  in which we vary
the spacing between SRRs along different lattice directions
independently. We observe large redshifts  and blueshifts in the
transmission resonances depending on SRR orientation relative to the
lattices. We establish that in-plane electric-electric dipole
coupling and out-of-plane magnetic-magnetic dipole coupling are
strong competing interactions. We explain the shifts by a
quasistatic electric and magnetic dipole coupling
model~\cite{Liu09a}, that enables us to determine the magnetic and
electric polarizability of SRRs. Finally, we discuss the role of
dynamic effects on the metamaterial resonance, which are evident in
density-dependent broadening and a saturation of the transmission.

\begin{figure}
\includegraphics[width=\figwidth]{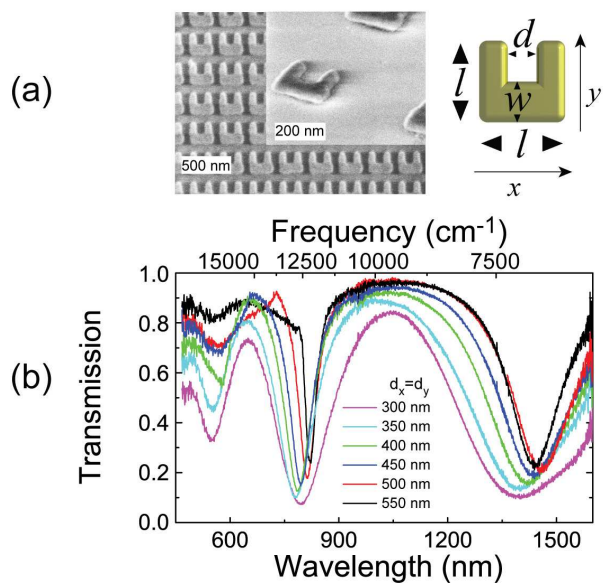}
\caption{(Color) (a) We fabricated arrays of Au SRRs on glass with
periodicities $d_{x,y} = 300$~nm (SEM micrograph) and larger
(inset). Each SRR has   $l=200$~nm,  $w=80$~nm, and SRR height
$30$~nm ($\pm 5$~nm). (b) Transmission spectra for square SRR arrays
with split width $d=80$~nm (polarization along $x$). The magnetic
resonance at 1.4~$\mu$m blue-shifts and broadens with increasing
density.} \label{fig1:SEM}
\end{figure}

We have fabricated Au SRRs on glass substrates by electron beam
lithography and lift-off using PMMA resist~\cite{Koenderink07},
without any adhesive layers. We took great care to produce SRRs of
identical dimension in arrays of different densities, using image
analysis of SEM micrographs (see Fig.~\ref{fig1:SEM}(a)) to overcome
proximity effects. Based on~\cite{Enkrich05}, our SRRs (200~nm base)
are expected to have an LC resonance at 1.4~$\mu$m. Although driven
by the electric field~\cite{Katsarakis04}, we refer to the resonance
as `magnetic', consistent with literature~\cite{Pendry01}.
 To
resolve the coupling strength between SRRs along the $x$ (along SRR
base) and $y$ (along SRR arms) directions separately, we varied the
pitches $d_x$ and $d_y$ independently between 300 nm and 550 nm,
staying below 550 nm to avoid grating diffraction in the range of
the magnetic resonance. We measured polarization-resolved normal
incidence transmission using the set up reported in
Ref.~\cite{Koenderink07}. We illuminated a mm-sized area on the
sample with a beam from a halogen lamp (5$^\circ$ opening angle) and
used a 20~$\mu$m pinhole in an intermediate image plane to  select
the transmitted intensity from single $36\times36~\mu$m$^2$ SRR
arrays, which we spectrally resolved by cooled Si CCD and InGaAs
array spectrometers, and normalized to transmitted intensity through
bare substrate.

\begin{figure}
\includegraphics[width=\figwidth]{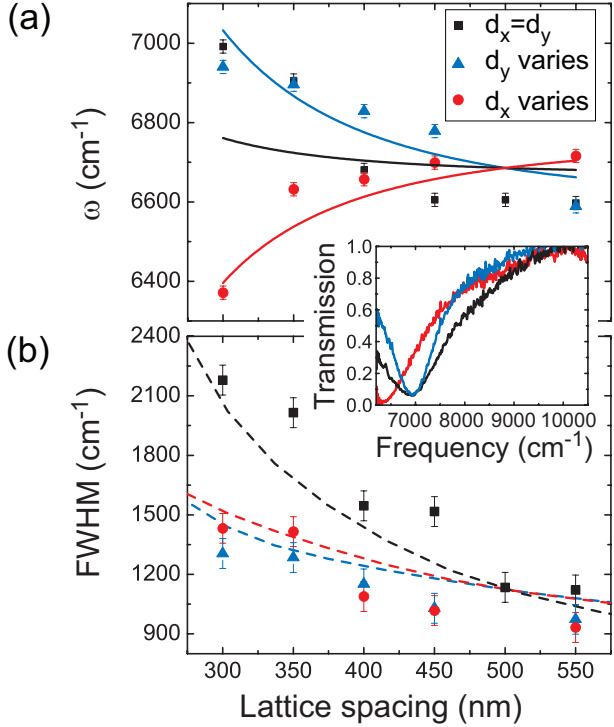}
\caption{(Color)(a) Frequency of the magnetic resonance versus
lattice spacing. The frequency blue-shifts when decreasing $d_y$
whether $d_y=d_x$ (black squares) or not (blue circles,
$d_x=500$~nm), while it red-shifts when decreasing $d_x$ (red
triangles, $d_y=500$~nm). The inset shows raw spectra for
$d_x=d_y=300$~nm (black curve), $d_x=500~$nm, $d_y=300$~nm (blue
curve), $d_x=300$~nm, $d_y=500$~nm (red curve). (b) FWHM of the
magnetic resonance versus lattice spacing (color coding as in (a)).
Curves are theory (electrostatic in (a), electrodynamic in (b)). }
\label{fig2:frequencies}
\end{figure}

Fig.~\ref{fig1:SEM}(b) shows $x$-polarized transmission spectra
measured on a sample with square lattices ($d_x=d_y$) of SRRs with
split width $d=80$~nm. We observe the magnetic resonance at
1.4~$\mu$m only for polarization along $x$, as reported by
\cite{Rockstuhl06,Enkrich05,Katsarakis04}, as well as higher order
plasmon resonances at~500 nm and 800 nm~\cite{Enkrich05}. In this
Letter we focus on the magnetic resonance. Tracing the minimum in
transmission  versus SRR density in Fig.~\ref{fig1:SEM}(b), we find
that the resonance blue-shifts as SRRs are brought closer. A
blueshift upon increased coupling is expected by analogy with
plasmon hybridization~\cite{Brongersma00,Prodan03,Koenderink06},
since the magnetic dipoles are all oriented perpendicular to the SRR
plane, and hence transversely coupled. To study this coupling in
detail, we fabricated samples with a large set of SRR arrays (split
width $d=100$~nm) where $d_x$ and $d_y$ are varied independently. We
expect a blueshift with increasing density for all arrays since the
magnetic dipoles are always transversely coupled. In
Fig.~\ref{fig2:frequencies}(a) we plot the measured center frequency
of the resonance versus SRR spacing for three sets of arrays. For
square lattices $d_x=d_y$ we indeed observe a continuous blue-shift,
confirming the data for  $d=80$~nm in Fig.~\ref{fig1:SEM}(b). We
also observe a blueshift when only $d_y$ is varied ($d_x=500$~nm).
Remarkably, we measure a redshift when only $d_x$ decreases and
$d_y$ is fixed at 500 nm. This result is surprising since red-shifts
imply longitudinal coupling, which is inconsistent with the
orientation of the magnetic dipoles. The redshift can only be
understood by noting that SRRs also have an electric polarizability
in addition to a magnetic dipole~\cite{Liu09a,Katsarakis04}. The
electric dipole moment points along the SRR base, hence allowing for
longitudinal coupling. We implement a model that takes into account
simultaneous electric and magnetic dipole coupling, similar to the
model  for  SRR stereodimers  reported in~\cite{Liu09a}. In this
model, all magnetic dipoles couple transversely while electric
dipoles transversely couple along $y$ and longitudinally along $x$.
We limit ourselves to electrostatic and magnetostatic
nearest-neighbor coupling, ignoring electro-dynamic effects, the
air-glass interface, and multipole corrections. However, this model
captures the main physics embodied in our observations.  The coupled
resonances are set by the Lagrangian
\begin{eqnarray}
{\cal{L}}=\sum_{i,j}\big[\frac{L}{2}(\dot{Q}_{i,j}^2-\omega_0^2
Q_{i,j}^2)-\frac{M_h}{d_x^3}\dot{Q}_{i,j}\dot{Q}_{i+1,j}
 \nonumber \\
- \frac{M_h}{d_y^3}\dot{Q}_{i,j}\dot{Q}_{i,j+1} +2\frac{M_e
\omega_0^2}{d_x^3}Q_{i,j}Q_{i+1,j}
 -\frac{M_e\omega_0^2}{d_y^3}Q_{i,j}Q_{i,j+1}\big], \nonumber \\
\label{Eq:Lagrangian}
\end{eqnarray}
where $L$ is the SRR inductance,  $Q_{i,j}(\dot{Q}_{i,j})$
represents the charge (current) on the SRR at  site $(i,j)$, and
where $M_h$ and $M_e$ quantify the mutual inductance and the
electric dipole coupling. Solving Eq.~(\ref{Eq:Lagrangian}) for the
resonance frequency at normal incidence ($k_\parallel=0$) yields
\begin{equation}
\omega
=\omega_0\sqrt{\frac{1-\frac{4\kappa_e}{d_x^3}+\frac{2\kappa_e}{d_y^3}}{1-\frac{2\kappa_h}{d_x^3}-\frac{2\kappa_h}{d_y^3}}}
\label{Eq: frequency}
\end{equation}
where $\omega_0$ is the resonance frequency of a single SRR, and
 $\kappa_{e,h}={M_{e,h}}/{L}$. Eq.~(\ref{Eq: frequency})
 is similar to a prediction by Marqu\'es et
al.~\cite{Marques08} for 3D SRR arrays. We fix $\kappa_e=1.04\cdot
10^{-21}$~m$^3$ to match the SRR electric polarizability to the
resonant extinction cross section of 0.3~$\mu$m$^2$ measured by
Husnik \emph{et al.}~\cite{Husnik08,polarnote}, and we set
$\kappa_h=0.67\kappa_e$.
\begin{figure}
\includegraphics[width=\figwidth]{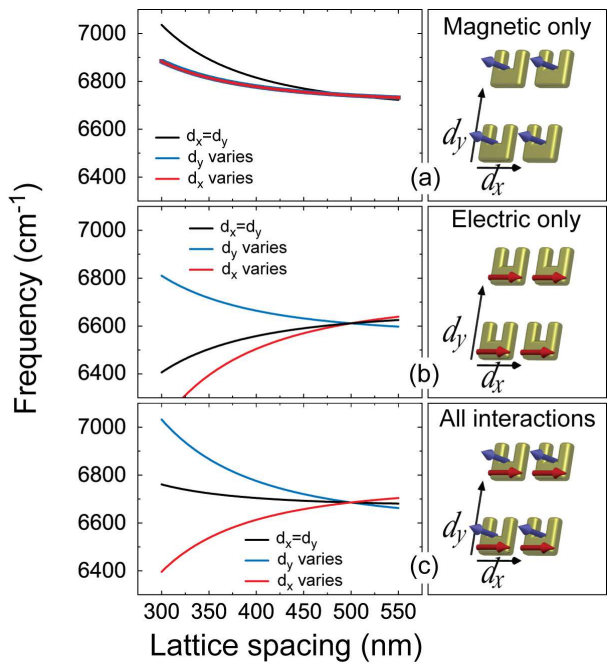}
\caption{(Color) Electrostatic calculation of the magnetic resonance
frequency as a function of lattice spacing. Black curves: $d_x=d_y$.
Blue curves: $d_y$ varies at fixed $d_x=500$~nm. Red curves: $d_x$
varies at fixed $d_y=500$~nm. For magnetic coupling only (a),
resonances always blue-shift with decreasing lattice spacing, while
for electric coupling only (b), the behavior of the resonances for
$d_x=d_y$  changes sign with respect to (c) (all couplings). Insets
in (a), (b) and (c) are sketches of the electric and magnetic
coupling between SRRs. Curves in (c) are reproduced in
Fig.~\ref{fig2:frequencies}.} \label{fig3:model}
\end{figure}
Fig.~\ref{fig3:model}(a) shows the resonance $\omega$ versus
$d_{x,y}$ assuming magnetic coupling only ($\kappa_e=0$). The
resonance blue-shifts for decreasing $d_{x,y}$ in all cases due to
transverse magnetic dipole coupling. Fig.~\ref{fig3:model}(b) shows
the resonance frequency for electric coupling only ($\kappa_h=0$).
The resonance red-shifts with increasing density unless $d_x$ is
fixed at 500~nm. This result indicates that longitudinal coupling
exceeds transverse coupling in square lattices of strictly in-plane
dipoles. This behavior is indeed observed for the purely electric
resonance at 800 nm, at least in the regime $\max(d_x,d_y)\leq
400$~nm where grating anomalies~\cite{Abajo07} (asymmetric shoulders
at 750 nm in Fig.~\ref{fig1:SEM}(b)) do not yet set in.

Neither model with solely electric or solely magnetic interaction is
consistent with the measured  shift of the 1.4$~\mu$m resonance,
since we observe blue-shifts in all cases except when $d_x$ is
varied and $d_y$ is fixed. Fig.~\ref{fig3:model}(c) shows the
calculated  $\omega$ taking into account both electric and magnetic
interactions. As in the data, the resonance only red-shifts when
decreasing the distance $d_x$ at large $d_y$. In this case
longitudinal electric coupling exceeds the sum of transverse
electric and magnetic coupling, leading to a net redshift. For a
quantitative comparison with our data we plot the shifts in
Fig.~\ref{fig3:model}(c) together with the data in
Fig.~\ref{fig2:frequencies}. The good quantitative agreement without
any adjustable parameters confirms our interpretation that SRRs in
metamaterial arrays show strong electric and magnetic dipole-dipole
interactions. These interactions are best quantified in rectangular
arrays, since   in square arrays studied
sofar~\cite{Enkrich05,Rockstuhl06} partial cancellation obscures the
magnetically induced blueshift. In our comparison we used $\kappa_e
/\kappa_h= 1.5$ measured in~\cite{Liu09a} for vertically stacked
SRRs. The data show that this ratio is also relevant for
dipole-dipole coupling in the $xy$-plane, allowing a direct
identification of $\kappa_e$ and $\kappa_h$ with on-resonance
electric and magnetic polarizabilities~(see note \cite{polarnote}).
It is  remarkable  that the magnetic polarizability $\alpha_M$ is of
the same order as the electric polarizability $\alpha_E$, as opposed
to the normal ordering $\alpha_M\ll\alpha_E$~\cite{Merlin09}. This
conclusion is in accordance with recent estimates of
Merlin~\cite{Merlin09}, that SRRs have $\alpha_M$ comparable in
magnitude to $\alpha_E$ provided $\mathrm{Im}
\epsilon_{\mathrm{Au}}\gg\lambda/\ell$, where $\ell$ is the
characteristic scatterer size. Given the dielectric constant of Au
$\mathrm{Im}\epsilon_{\mathrm{Au}} \sim 10$,  and the size
$\lambda/\ell\sim 7$ of our SRRs, their LC resonances  are indeed
expected to be magnetic resonances with large $\alpha_M$. Into the
visible, $\mathrm{Im} \epsilon_{\mathrm{Au}}$ rapidly decreases,
causing $\alpha_M$ to vanish~\cite{Merlin09}, as argued
independently in~\cite{Klein06}.

A striking feature in our transmission data (\emph{cf.}
Fig.~\ref{fig1:SEM}(b)) in addition to the spectral shifts, is the
large  broadening of the resonance as the density of SRRs increases.
In Fig.~\ref{fig2:frequencies}(b) we plot the measured full width at
half minimum (FWHM) of the transmission minimum versus lattice
spacing. For square lattices, the width more than doubles from 1000
to 2150~cm$^{-1}$ as the pitch is reduced from 550 to 300 nm, while
for both types of rectangular lattices ($d_x$ or $d_y$ fixed at 500
nm) the width increases  from 950 to 1400~cm$^{-1}$. Such broadening
was also noted  by Rockstuhl et al.~\cite{Rockstuhl06} for square
arrays. Our extensive data on many rectangular and square arrays
allow us to quantitatively identify the source of broadening. From
the outset it is clear that the broadening is outside the scope of
Eq.~(\ref{Eq:Lagrangian}), since the (Ohmic) damping rate is almost
independent of coupling in  any electrostatic model. Instead,
electrodynamical radiation damping, i.e., scattering loss into the
far field must be taken into account. As all oscillators in our
sub-diffraction lattices are driven in phase ($k_{||}=0$), scattered
light radiated by all oscillators interferes destructively for all
angles, except along the transmitted and reflected direction. Since
the magnetic dipoles are aligned along the incident beam, they do
not radiate any amplitude into the $k_{||}=0$ directions. Hence, all
radiation damping is \emph{solely} due to the induced
\emph{electric} dipoles. For a quantitative analysis we use an
electrodynamical model for  electric point dipoles with a Lorentzian
resonance in $\alpha_E$ according
to~\cite{polarnote,Koenderink06,Abajo07}, centered at 1.4~$\mu$m and
including the material loss rate of Au, in addition to radiation
damping. We evaluate Eqs.~(8,9) in Ref.~\cite{Abajo07} to predict
the array transmission. This dynamic model has no adjustable
parameters, since the on-resonance polarizability is
fixed~\cite{polarnote} to match the extinction cross section of
single SRRs in~\cite{Husnik08}.  We find a broadening of the
collective transmission resonance that quantitatively reproduces the
measured broadening with decreasing pitch for all lattices (FWHM
curves in Fig.~\ref{fig2:frequencies}(b)). An important conclusion
is that the large width of the magnetic response commonly observed
for SRR
arrays~\cite{Smith00,Enkrich05,
Rockstuhl06} is not due to
intrinsic loss, but is quantitatively consistent with superradiant
decay of the electric dipoles. The collective enhancement of the
single SRR radiative linewidth, already suspected
by~\cite{Rockstuhl06}, implies enhanced scattering and  a reduction
of the absorption of the array far below the albedo of single SRRs.
\begin{figure}
\includegraphics [width=\figwidth]{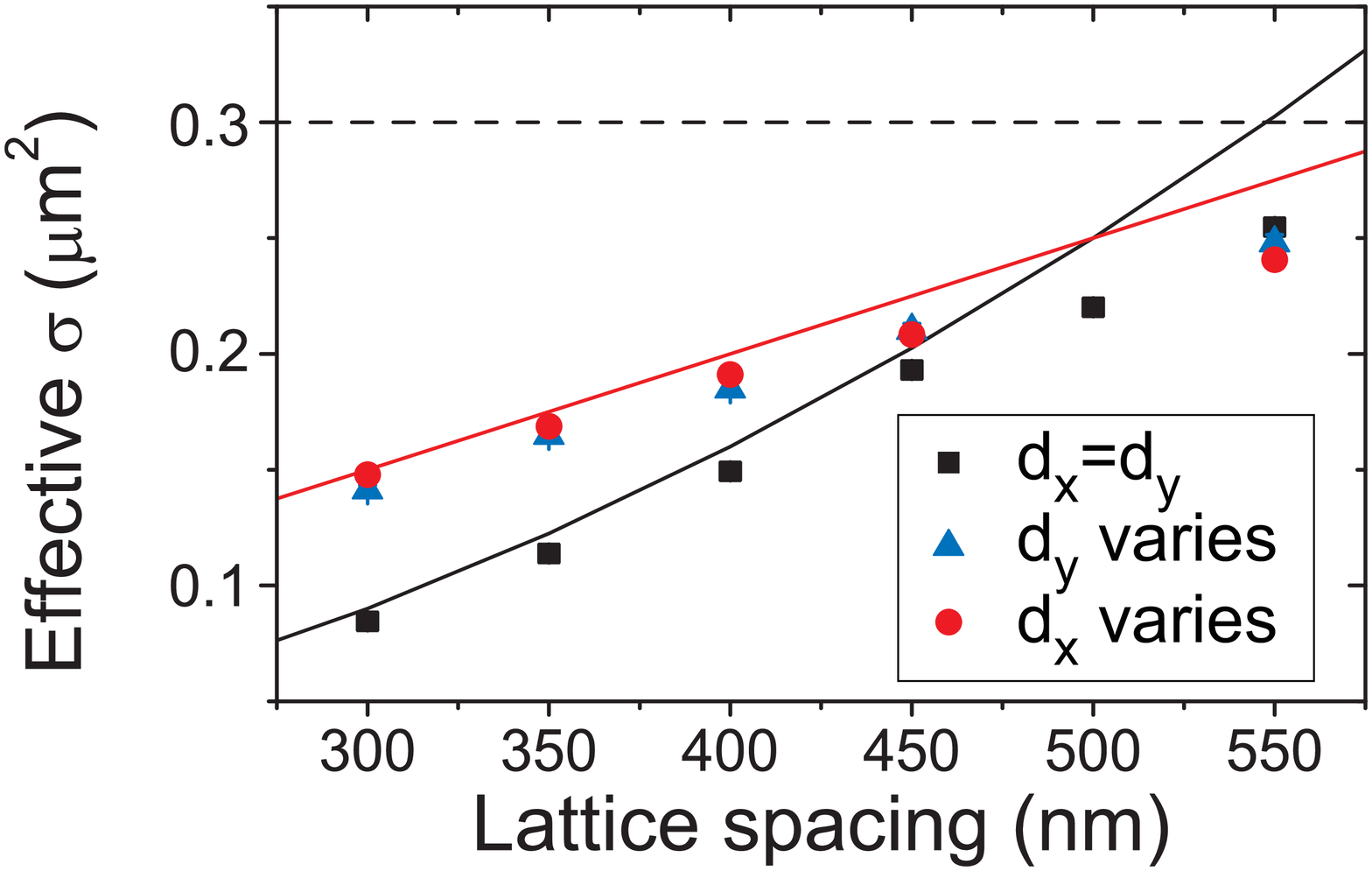}
\caption{Effective extinction cross section per SRR  derived from
on-resonance tranmission. The black dashed line indicates the cross
section of a single SRR (from~\cite{Husnik08}). The cross section
per SRR is limited by the area $d_xd_y$ of the unit cell (black and
red lines).} \label{fig4:extinctiondip}
\end{figure}

Finally we correlate the resonance broadening with the measured
transmission $T$ on resonance. Fig.~\ref{fig4:extinctiondip} shows
the effective extinction cross section derived from our measurements
through $\sigma_{\text{eff}}=d_xd_y(1-T)$. For uncoupled scatterers
 we expect constant $\sigma_{\text{eff}}$ equal to
the extinction cross section $\sigma_{\mathrm{ext}}=0.3~\mu$m$^2$
measured for a single SRR in~\cite{Husnik08} (dashed line in
Fig.~\ref{fig4:extinctiondip}), as indeed almost found in our data
for $d_x=d_y
>500$~nm.  For $d<500$~nm, we measure values for $\sigma_{\text{eff}}$ far
below $\sigma_{\mathrm{ext}}$ indicative of strong dipole-dipole
coupling. The collective superradiant decay
(Fig.~\ref{fig2:frequencies}(b)) which widens the resonance  reduces
the extinction per element to remain below the unit-cell area
$d_xd_y$ (curves in Fig.~\ref{fig4:extinctiondip}).

In conclusion, we have measured large resonance shifts as a function
of density in SRR arrays resonant at $\lambda=1.4~\mu$m. These
shifts  are due to strong near-field electrostatic and magnetostatic
dipole coupling. Furthermore, we  observe electrodynamic
superradiant damping that causes resonance broadening and an
effective reduction of the extinction cross section per SRR. Since
the data show that the response of SRR arrays is not simply given by
the product of the density  and
 polarizability of single constituents, we conclude that  a
Lorentz-Lorenz analysis to explain effective media parameters of
metamaterials `atomistically' is not valid~\cite{Simovski07}. The
fact that the Lorentz-Lorenz picture is invalid has important
repercussions: It calls for a shift away from the paradigm that the
highest polarizability per constituent is required to obtain the
strongest electric or magnetic response from arrays of electric or
magnetic scatterers. Our experiments show that increasing the
density of highly polarizable constituents to raise the effective
medium response~\cite{Enkrich05} is ineffective, since superradiant
damping limits the achievable response. To strengthen $\epsilon$ or
$\mu$, we propose that one ideally finds constituents that have both
a smaller footprint \emph{and a smaller polarizability} per
constituent. We stress that even if constituent coupling modifies
$\epsilon$ and $\mu$, we do not call into question reported
effective medium parameters or the conceptual validity thereof per
se. The effective medium regime only breaks down when constituent
coupling is so strong that collective modes of differently shaped
macroscopic objects carved from the same SRR array have very
different resonance frequencies or widths. In this regime
interesting physics comes into view, particularly regarding active
devices. Specific examples are array antennas for spontaneous
emission~\cite{Li06} and `lasing spasers'~\cite{Zheludev08}, where
the lowest-loss array mode will lase most easily.

\begin{acknowledgments}
We thank Chris R\'etif, Dries van Oosten, Jaime G\'{o}mez Rivas,
Albert Polman and Kobus Kuipers for assistance and discussions. This
work is part of the research program of the ``Stichting voor
Fundamenteel Onderzoek der Materie (FOM),'' which is financially
supported by the ``Nederlandse Organisatie voor Wetenschappelijk
Onderzoek (NWO).''
\end{acknowledgments}

\end{document}